\documentclass[fleqn,twoside]{article}
\usepackage{amsmath,amssymb,amsbsy,mathbbol}
\usepackage{graphicx,subfigure,psfrag}
\usepackage{espcrc2}

\title{Applying chiral perturbation to twisted mass lattice
  QCD\thanks{Partially supported by the U.S. Department of Energy
    Grant No. DE-FG02-96ER40956}\thanks{Talk presented by J. Wu at
    Lattice 2004, Fermi National Accelerator Laboratory, June 21 - 26,
    2004.}}

\author{Stephen R. Sharpe\address[UW]{Physics Department, Box 351560,
    University of Washington, Seattle, Washington 98195, USA} and
    Jackson M. S. Wu\addressmark[UW]}
       
\begin{document}

\begin{abstract}
We have explored twisted mass LQCD (tmLQCD) analytically using chiral
perturbation theory, including discretization effects up to $O(a^2)$,
and working at next-to-leading (NLO) order in the chiral expansion. In
particular we have studied the vacuum structure, and calculated the
dependence of pion masses and decay constants on the quark mass,
twisting angle and lattice spacing. We give explicit examples for
quantities that both are and are not automatically improved at maximal
twisting. 
\vspace{1pc}
\end{abstract}

\maketitle

\section{Introduction}
There has been much interest recently in the twisted mass formulation
of lattice QCD~\cite{Frezetal,FR03} (see Ref.~\cite{FrezLat04} for a
recent review). We study here the importance of the symmetry breaking
inherent in tmLQCD, which can be done analytically using the chiral
effective theory including effects of discretization.

\section{\label{sec:ECT}The effective chiral Lagrangian}
To derive an effective continuum chiral theory for tmLQCD, we follow
the two-step procedure of Ref.~\cite{SS98}. We first write down
an effective continuum Lagrangian at the quark level that describes
the long distance physics of the underlying lattice theory, and then
match it onto an effective chiral Lagrangian. The details of this
procedure are given in Ref.~\cite{SW04}. 

The effective continuum Lagrangian is constrained to be invariant
under the symmetries of the lattice theory. It has form\footnote{The
  $O(a^2)$ terms do not contribute to any further symmetry breaking
  than the terms explicity shown.}: 
\setlength\arraycolsep{2pt}
\begin{eqnarray} \label{E:CLeff}
\mathcal{L}_{eff} = \mathcal{L}_g &+& \bar{\psi}(D \!\!\!\! / + m 
+ i \gamma_5 \tau_3 \mu) \psi \notag \\
&+& b_1 a \bar{\psi} i \sigma_{\mu\nu} F_{\mu\nu} \psi +
O(a^2) \,, 
\end{eqnarray}
where $\mathcal{L}_g$ is the continuum gluon Lagrangian, $m$ is the
physical quark mass, defined as usual by $m = Z_m(m_0 - \widetilde
m_c)/a$, and $\mu$ is the physical twisted mass defined by $\mu =
Z_\mu \mu_0/a$ ($m_0$ and $\mu_0$ are bare normal and twisted mass 
parameters respectively). Note that the lattice symmetries forbid
additive renormalization of $\mu_0$~\cite{Frezetal}.

In writing down (\ref{E:CLeff}), we have dropped terms that vanish by
equations of motion. We have also dropped terms of order higher than
quadratic in $\{m,\mu,a\}$ (factors of $\Lambda_{QCD}$ are implicit
here and henceforth) in anticipation of the power-counting in our
chiral effective theory. The net result is that (\ref{E:CLeff})
differs from that for the standard Wilson theory given in~\cite{SS98}
only by the addition of a twisted mass term. 

Next we write down a generalization of the continuum chiral Lagrangian
that includes the effects of the Pauli term. We use the power counting
scheme: 
\begin{equation} \label{AokiPC}
1 \gg \{\mathsf{m},p^2, a\} \gg
\{\mathsf{m}^2,\mathsf{m}p^2,p^4,a\mathsf{m},ap^2,a^2\} \,,
\end{equation}
where \textsf{m} is a generic mass parameter that can be either $m$ or
$\mu$. The chiral Lagrangian is built from the standard $SU(2)$
matrix-valued field $\Sigma$. 
It can be obtained from the quark Lagrangian, (\ref{E:CLeff}), by a
standard spurion analysis. We must introduce a spurion matrix $\hat A$
for the Pauli term, as well as the usual spurion $\chi$ for the mass 
terms. Both transform in the same way as the $\Sigma$ field. At the
end of the analysis the spurions are set to their respective constant
values:
\setlength\arraycolsep{2pt}
\begin{eqnarray}
\chi &\rightarrow& 2 B_0 (m + i \tau^3 \mu) \equiv \hat{m} 
+ i \tau^3 \hat{\mu} \,, \notag \\
\hat A &\rightarrow& 2 W_0 \, a \equiv \hat{a} \,,
\end{eqnarray}
where $B_0$ and $W_0$ are unknown dimensionful parameters. The
difference from the standard construction comes only from the
inclusion of the $\mu$ term in the constant value of $\chi$. Thus, we
can read off the form of the chiral Lagrangian for tmLQCD from
Ref.~\cite{ORS03}, where the Lagrangian for untwisted Wilson fermions
was worked out to quadratic order in our expansion\footnote{Similar
  work has been done in Ref.~\cite{MS04}. We have extended that work
  by including $O(a^2)$ terms.}. The only extension we make is to
include sources for currents and densities. 

\section{\label{sec:tmLphase}The phase diagram of tmLQCD}
Using the chiral Lagrangian, we have extended the analysis of
Ref.~\cite{SS98} into the twisted mass plane. The details of the
analysis are given in Ref.~\cite{SW04}. We record here the main
results.\footnote{Similar considerations have been made in
  Refs.~\cite{Mun04,Scor04}}

In the region $a \gg m' \sim \mu \sim a^2$ where the Aoki phase occurs
for untwisted Wilson fermions ($m' = m + aW_0/B_0$ is the shifted mass
due to effect of the Pauli term, as noted in Ref.~\cite{SS98}), the
potential energy is dominated by:
\setlength\arraycolsep{2pt}
\begin{eqnarray}
V_\chi &=& -\frac{c_1}{4} \mathrm{Tr}(\Sigma + \Sigma^{\dagger}) 
+ \frac{c_3}{4} \mathrm{Tr} \big[i(\Sigma - \Sigma^{\dagger}) 
\tau_3 \big] \notag \\
&&+ \frac{c_2}{16} \left[\mathrm{Tr}
(\Sigma + \Sigma^{\dagger})\right]^2 \,, 
\end{eqnarray}
where $c_1 \sim m'$, $c_3 \sim \mu$, and $c_2 \sim a^2$ are all of
the same order. As in the untwisted case, it is the competition
between the leading order mass terms and the $a^2$ term at NLO that
can lead to interesting phase structure.

To determine the vacuum structure we find the condensate, $\Sigma_0$,
which minimizes the potential. Parametrizing using $\Sigma_0 = A_m +
i\mathbf{B}_m \cdot \boldsymbol{\tau}$, where $A_m^2 + \mathbf{B}_m^2
= 1$, the functional forms of $A_m$ and $\boldsymbol{B}_m$ are found
by solving a quartic equation. The minimum is found to lie in the
$\mathbb{1}$-$\tau_3$ plane in flavor space.

\begin{figure}[t]
\centering
\subfigure[Phase diagram for $c_2 < 0$]{
\psfrag{b}[][ct]{\small $c_3/c_2 \sim \mu/a^2$}
\psfrag{a}{\large $\frac{c_1}{c_2} \sim \frac{m'}{a^2}$}
\label{fig:mmu:a}
\includegraphics[width=2.1in]{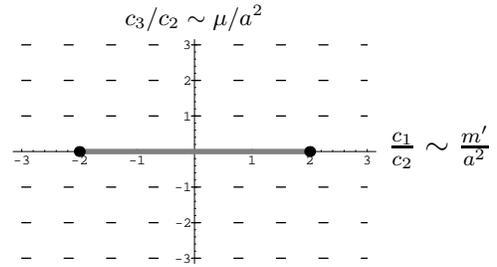}}
\vspace{0.1in}
\subfigure[Phase diagram for $c_2 < 0$]{
\psfrag{b}[][ct]{\small $c_3/c_2 \sim \mu/a^2$}
\psfrag{a}{\large $\frac{c_1}{c_2} \sim \frac{m'}{a^2}$}
\label{fig:mmu:b}
\includegraphics[width=2.1in]{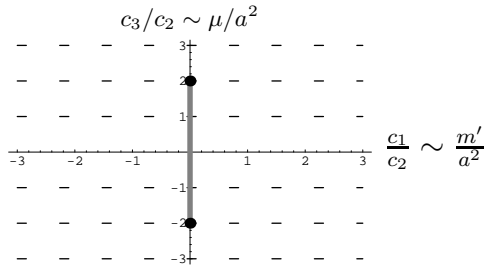}}
\vspace{-0.3in}
\caption{\label{fig:mmu} Phase diagram of tmLQCD. The sign of the
  coefficient $c_2$ determines whether flavor symmetry is
  spontaneously broken in the standard Wilson theory. The solid lines
  are first order transitions across which the condensate is
  discontinuous, with second-order endpoints.}
\vspace{-0.3in}
\end{figure}

The vacuum phase structure depends on the sign of the coefficient
$c_2$, which is a linear combination of low-energy constants (LEC's)
in the chiral Lagrangian. Fig.~\ref{fig:mmu} illustrates the two
possibilities.

For $c_2 > 0$, the usual Aoki phase (first noted in
Ref.~\cite{Aoki84}) occurs, but is confined to a short segment on the
untwisted axis. In the Aoki phase segment, the charged pions are
massless but the neutral pion is not. All three pions become massless
only at the second-order endpoints.

For $c_2 < 0$, there is a first-order transition line extending a
distance $\Delta \mu \sim a^2$ in the twisted direction. The charged
pions are not massless anywhere on this first-order line (including
the endpoints), while the neutral pion becomes massless at the two
second-order endpoints. In fact the neutral pion mass for $c_2 < 0$
has the same depenence on $\mu$ along the first-order transition line
as that on $m$ along the Aoki phase line for $c_2 > 0$. Evidence for
the phase structure in the $c_2 < 0$ scenario has recently been seen
in numerical simulations~\cite{Faretal}.

\section{\label{sec:MpiFpi}Pion masses and pion decay constant}
In the region $m \sim \mu \sim a \gg a^2 \sim a\mathsf{m} \sim
\mathsf{m}^2$ away from the ``Aoki phase'' (where most simulations
will be done), we find that the splitting of pion masses due to
parity-flavor breaking first occurs at NLO: 
\begin{equation} \label{E:msplit}
\Delta m_\pi^2 \equiv m_{\pi_3}^2 - m_{\pi_{1,2}}^2 =
\frac{\mathcal{C}\,\hat{a}^2 \hat{\mu}^2}
     {(\hat{m} + \hat{a})^2 + \hat{\mu}^2} \,,  
\end{equation}
where $\mathcal{C}$ is a linear combination of the LEC's. The mass
splitting is $\propto a^2$ for any values of $\hat{m}$ and $\hat{\mu}$
within the region we are considering, not just at maximal
twisting. For $\hat{m}, \hat{\mu} \gg \hat{a}$, (\ref{E:msplit}) 
becomes $\mathcal{C}\, \hat{a}^2 \sin^2 \omega$, the form expected
from the symmetries of tmLQCD~\cite{FR03}. Note that at NLO, there are
only tree level contributions to the mass splitting, whose full
expression agrees with that of Ref.~\cite{Scor04}. For pion decay
constants, we find no splitting until next-to-next-to-leading order.

The results of the one-loop calculation of the pions masses, the
wavefunction renormalizations, and the pion decay constants will be
given elsewhere~\cite{SWprep}.  

\section{\label{sec:MtxElt}Current matrix elements}
Once the twisting angle is defined, the physical untwisted currents
can be obtained simply from the appropriate functional derivative of
the effective chiral Lagrangian with respect to the sources. We use the
non-perturbative definition given by:
\begin{equation}
\tan \omega = \langle \partial_\mu V_\mu^2 P^1 \rangle / 
              \langle \partial_\mu A_\mu^1 P^1 \rangle \,,
\end{equation}
where the defining correlators are evaluated in the chiral theory
using single pion contributions. We find the twisting angle is related
to the parameters in the chiral theory by: 
\begin{equation}
\tan \omega = B_{3,m}/\big( A_m + 4W_{10}\hat{a}/f^2 \big) \,, 
\end{equation}
where $W_{10}$ is one of the unknown LEC's.

As an example, consider the physical untwisted axial vector current:
\setlength\arraycolsep{2pt}
\begin{eqnarray}
[A^a_\mu](\omega) &=& \big( \cos (\omega) (1-\delta^{3a}) +
\delta^{3a} \big) A_\mu^a \notag \\
&& + \sin (\omega) \varepsilon^{3ab} V_\mu^b \,.
\end{eqnarray}
The effect of parity-flavor breaking can be seen when $[A^a_\mu]$ is
expanded in terms of the pion fields. There are terms with even number
of pion fields that are usually forbidden by parity, and the
coefficients of the expansion depend on the flavor index. 

The physical, single pion matrix element of the axial vector current,
$\langle 0 | [A_\mu^a] | \pi_b \rangle$, which defines the pion decay
constant, is an example of a quantity that is automatically $O(a)$
improved. But the unphysical, two-pion matrix elements are not (and
are not expected to be, according to the general theory of
Ref.~\cite{FR03}). For instance, let $q = p_2 - p_1 \ll m_\pi^2$, then
for $a=1,2$, we find:
\setlength\arraycolsep{2pt}
\begin{eqnarray}
\lefteqn{\langle \pi_a(p_1) | [A_\mu^a] | \pi_3(p_2) \rangle =
- i (p_1 + p_2) \frac{B_{3,m} \hat{a}}{f^2}\mathcal{D}_1} \notag \\
&&\; + i q \Big[ \frac{4 B_{3,m} \hat{a}}{f^2}\mathcal{D}_2  
- \frac{8 B_{3,m}}{f^2}\big(\hat{a}\mathcal{D}_3 
+ \frac{\hat{a}^3}{m_\pi^2}\mathcal{D}_4\big) \Big] \,, 
\end{eqnarray}
where the $\mathcal{D}_i$'s are linear combinations of LEC's. The
$1/m_\pi^2$ term comes from a diagram which contains a three-pion
vertex that arises from parity-flavor breaking. Since the continuum
limit must be taken before the chiral limit to avoid entering the Aoki
phase, there is no divergence. 

A potential use of the unphysical two-pion current matrix elements is
to provide a way of calculating the new LEC's introduced by
discretization into the chiral Lagrangian.

\end{document}